\documentclass[english]{iopart}
\usepackage[T1]{fontenc}
\usepackage[latin9]{inputenc}
\usepackage{babel}
\usepackage{amstext}
\usepackage{amssymb}
\usepackage{graphicx}
\usepackage[unicode=true,
 bookmarks=true,bookmarksnumbered=false,bookmarksopen=false,
 breaklinks=false,pdfborder={0 0 1},backref=false,colorlinks=false]
 {hyperref}

\makeatletter
\usepackage{iopams}
\usepackage{setstack}

\makeatother

\begin{document}

\title{Dissipative optomechanical squeezing of light}

\author{Andreas Kronwald$^1$, Florian Marquardt$^{1,2}$, and Aashish A.
Clerk$^3$}

\address{$^1$Friedrich-Alexander-Universität Erlangen-Nürnberg, Staudtstr.
7, D-91058 Erlangen, Germany}

\address{$^2$Max Planck Institute for the Science of Light, Günther-Scharowsky-Straße
1/Bau 24, D-91058 Erlangen, Germany}

\address{$^3$Department of Physics, McGill University, Montreal, Quebec,
Canada H3A 2T8 }

\ead{andreas.kronwald@physik.uni-erlangen.de}

\begin{abstract}
We discuss a simple yet surprisingly effective mechanism which allows
the generation of squeezed output light from an optomechanical cavity.
In contrast to the well known mechanism of \textquotedbl{}ponderomotive
squeezing\textquotedbl{}, our scheme generates squeezed output light
by explicitly using the dissipative nature of the mechanical resonator.
We show that our scheme has many advantages over ponderomotive squeezing;
in particular, it is far more effective in the good cavity limit commonly
used in experiments. Furthermore, the squeezing generated in our approach
can be directly used to enhance the intrinsic measurement sensitivity
of the optomechanical cavity; one does not have to feed the squeezed
light into a separate measurement device. As our scheme is very general,
it could also e.g.~be implemented using superconducting circuits.
\end{abstract}
\maketitle
\global\long\def\i{i}
\global\long\def\G{\mathcal{G}}
\global\long\def\a{\hat{a}}
\global\long\def\ad{\hat{a}^{\dagger}}
\global\long\def\d{\hat{d}}
\global\long\def\dd{\hat{d}^{\dagger}}
\global\long\def\b{\hat{b}}
\global\long\def\bd{\hat{b}^{\dagger}}
\global\long\def\be{\hat{\beta}}
\global\long\def\bed{\hat{\beta}^{\dagger}}
\global\long\def\C{\mathcal{C}}
\global\long\def\nth{n_{\text{th}}}
\global\long\def\neff{n_{\text{eff}}}
\global\long\def\eps{\varepsilon}
\global\long\def\x{\hat{x}}
\global\long\def\p{\hat{p}}
\global\long\def\Xone{\hat{X_{1}}}
\global\long\def\Xtwo{\hat{X_{2}}}
\global\long\def\sz{\sigma_{z}}

\section{Introduction}

Among the simplest kinds of non-classical light is squeezed light,
where fluctuations in one quadrature of the optical amplitude drop
below the level of vacuum noise. Such light is interesting from both
fundamental and practical points of view. Squeezed light can be used
to improve the measurement sensitivity in applications ranging from
gravitational wave detection \cite{1980_Caves_ForceMsmntQuantumMechOsc,2011_LIGO_squeezedlight_in_grav_wave_detector,2013_LIGO_Squeezed_light_impro_grav_wave_det}
to even biology \cite{2013_Taylor_SqueezedLightBiologicalSystem}.
Squeezed states of light are also a key ingredient for continuous-variable
information processing \cite{2005_BraunsteinRMP}. 

While the standard method for generating optical squeezing is to drive
a nonlinear optical medium (see, e.g.~\cite{2010_Eberle_Squeezing_Larger_10dB}),
it has long been realized \cite{1967_Braginsky_PonderomotiveEffects}
that squeezing can also be realized in optomechanical systems \cite{2013_AspelmeyerKippenbergMarquardt_OptomechanicsReview,2013_Meystre_OptomechanicsReview},
where cavity photons are coupled to mechanical motion by radiation
pressure. The standard mechanism for such squeezing, termed {}``ponderomotive
squeezing'' (PS) \cite{1967_Braginsky_PonderomotiveEffects}, relies
on the mechanical resonator effectively mediating a (coherent) Kerr-type
($\chi_{3})$ optical nonlinearity \cite{1994_Fabre_QuNoise,1997_Mancini_PonderomotiveControl};
as in a Kerr medium, squeezing is produced by generating classical
correlations between the amplitude and phase quadratures of light
leaving the cavity. This sort of ponderomotive squeezing has recently
been realized in experiments \cite{2012_Brooks_Non-classicalLight,2013_Safavi_Naeini_PonderomotiveSqueezing,2013_Purdy_PonderomotiveSqueezing}.
\begin{figure}[b]
\begin{centering}
\includegraphics[scale=1.25]{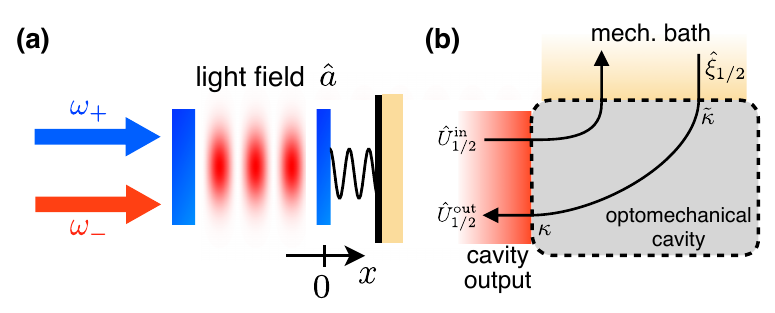}
\par\end{centering}

\caption{(a) Basic setup for dissipative generation of squeezed output light:
an optomechanical cavity driven by two lasers on the red and blue
mechanical sideband. By carefully tuning the amplitudes of the lasers,
strong squeezing is possible. (b) Schematic showing the basic idea
of the dissipative squeezing mechanism. Optical vacuum fluctuations
(red) entering the optomechanical cavity at a rate $\kappa$ are perfectly
absorbed by the mechanical resonator. At the same time, the damped
mechanical resonator acts as an effective squeezed dissipative bath
for cavity photons (even though the mechanical resonator itself is
in a \textit{thermal} state). The effectively squeezed mechanical
noise $\hat{\xi}$ (coupled to the cavity at a rate $\tilde{\kappa}$)
is optimally forwarded to the cavity output port $\hat{U}_{\text{out}}$,
i.e., the output light is maximally squeezed if the cavity decay rate
$\kappa$ equals $\tilde{\kappa}$.}

\label{FIG_1}
\end{figure}

In this work, we describe a fundamentally different and potentially
powerful new method for generating squeezed light using optomechanics,
cf.~Fig.~\ref{FIG_1}a. Unlike standard ponderomotive squeezing,
our scheme is not based on having the mechanics mediate a coherent
(i.e.,~Hamiltonian) optical nonlinearity; instead, it uses the dissipative
nature of the mechanical resonator. As we show, by using a (classical)
bichromatic cavity drive, the mechanics can be made to mimic a dissipative
squeezed reservoir. By careful tuning of the cavity laser drives,
this effective mechanical reservoir acts as a ``sink'' for the fluctuations
of the incident light, and imprints its squeezed noise almost perfectly
onto the output light (cf.~Fig.~\ref{FIG_1}b). We also show that
the squeezing generated in our approach can directly be used to enhance
the intrinsic measurement sensitivity of the optomechanical cavity
(i.e., to detect a signal coupled dispersively to the cavity). Note
that although we focus on an optomechanical implementation of our
scheme here, we stress that it could also be implemented using superconducting
circuits \cite{2009_Kamla_JosephsonParamAmpl,2010_Bergeal_JosephseonRingModulator,2013_Peropadre_TunableCouplingSupercondResonators}
as our scheme relies only on having two modes coupled parametrically
with both, beam-splitter and non-degenerate parametric amplifier terms.

Our scheme is well within the reach of current state-of-the-art optomechanical
experiments, some of which have already made use of two-tone driving
\cite{2010_Hertzberg_BackactionEvadingMeasurements,2010_WeisKippenberg_OMIT,2011_Teufel_StrongCouplingEIT,2011Safavi-Naeini:OMIT}.
As we discuss, it has several advantages over standard ponderomotive
squeezing. In particular, our scheme is efficient in the good-cavity
limit commonly used in experiments, and squeezes the same quadrature
of light over an appreciable bandwidth. This is to be contrasted against
PS, which is not efficient in the good-cavity limit, and produces
squeezing with a frequency-dependent squeezing angle. In addition,
the squeezing generated in our scheme can be used directly to enhance
cavity based measurements; one does not need to feed the squeezed
light into a separate measurement device (see Sec.~\ref{sec:Increasing-the-measurement}). 

Note that the scheme we describe is related to the protocol described
and implemented by Wasilewski et al.~\cite{Wasilewski:09} to generate
pulses of two-mode squeezed light. Their approach did not use mechanical
interactions, but rather interactions with two polarized atomic spin-ensembles,
each of which acts as an oscillator. While similar in spirit, there
are some important differences: our scheme generates continuous-wave
squeezed light, and makes use of dissipation in a fundamental way
(in contrast, Ref.~\cite{Wasilewski:09} does not treat atomic dissipation
as it plays no role in their approach). Our scheme is also related
to our earlier proposal for generating strong mechanical squeezing
in an optomechanical cavity \cite{2013_Kronwald_MechanicalSqueezing}
(which in turn is related to \cite{2014_Didier_Squeezing} and earlier
proposals \cite{1993_Cirac_DarkSqueezedStates,2004_Rabl_NanomechSqueezedStates,2006_Parkins_DissipativeTwoModeSqueezing,2013_DallaTorre_SteadyStateSqueezingSpinEnsemble}).
Unlike that problem, the interest here is on generating squeezing
of an output field (as opposed to an intracavity field); similar to
the situation with squeezing via parametric processes \cite{1984_Collett_Squeezing,1990_GeaBanacloche_OutputLightSqueezing_Not_IntracavityLightSqueezing},
there are crucial differences between these two goals.

\section{Model}

We consider a standard, single-sided optomechanical cavity, where
electromagnetic radiation couples to mechanical motion via radiation
pressure, cf.~Fig.~\ref{FIG_1}a (non-ideal or two-sided cavities
are discussed in the appendix). The optomechanical Hamiltonian reads
\cite{1995_Law_OptomechHamiltonian}
\begin{equation}
\hat{H}=\hbar\omega_{\text{cav}}\ad\a+\hbar\Omega\bd\b-\hbar g_{0}\left(\bd+\b\right)\ad\a+\hat{H}_{\text{dr}}\,.\label{eq:OM_Hamiltonian}
\end{equation}
where $\omega_{\text{cav}}$ ($\Omega$) is the cavity (mechanical)
resonance frequency, $\a$ ($\b$) the photon (phonon) annihilation
operator and $g_{0}$ the optomechanical coupling strength. $\hat{H}_{\text{dr}}=\hbar\left(\alpha\left(t\right)\ad+\text{h.c.}\right)$
is the coherent laser driving Hamiltonian where $\alpha\left(t\right)$
describes a general, coherent multi-tone laser drive. In the following,
we decompose the photon annihilation operator $\a=\bar{a}+\d$ into
a classical amplitude $\bar{a}$ and quantum fluctuations $\d$. Treating
cavity dissipation via standard input-output theory \cite{2008_ClerkDevoretGirvinFMSchoelkopf_RMP},
the dynamics of the quantum fluctuations is given by the quantum Langevin
equation 
\begin{equation}
\dot{\d}=\frac{\i}{\hbar}\left[\hat{H},\d\right]-\frac{\kappa}{2}\d-\sqrt{\kappa}\d_{\text{in}}\label{eq:langevin}
\end{equation}
where $\kappa$ is the cavity decay rate. The equation of motion for
the mechanical operator $\b$ reads
\[
\dot{\hat{b}}=\frac{\i}{\hbar}\left[\hat{H},\hat{b}\right]-\frac{\Gamma_{M}}{2}\hat{b}-\sqrt{\Gamma_{M}}\hat{b}_{\text{in}}\,,
\]
where $\Gamma_{M}$ is the mechanical decay rate. The non-zero noise
correlators read $\langle\d_{\text{in}}\left(t\right)\dd_{\text{in}}\left(t'\right)\rangle=\delta\left(t-t'\right)$,
$\langle\b_{\text{in}}\left(t\right)\bd_{\text{in}}\left(t'\right)\rangle=\left(\nth+1\right)\delta\left(t-t'\right)$
and $\langle\bd_{\text{in}}\left(t\right)\b_{\text{in}}\left(t'\right)\rangle=\nth\delta\left(t-t'\right)$,
where $\nth$ is thermal occupancy of the mechanical reservoir. 

Our interest is on the noise properties of the light leaving the cavity.
The fluctuations in the output light is described by $\d_{\text{out}}$,
which in turn is determined by the incident noise $\d_{\text{in}}$
and the intracavity light $\d$ via the input-output relation $\d_{\text{out}}=\d_{\text{in}}+\sqrt{\kappa}\d$
\cite{2008_ClerkDevoretGirvinFMSchoelkopf_RMP}. A general quadrature
of the output light is defined by 
\begin{equation}
\hat{U}_{\varphi}^{\text{out}}=\left(\d_{\text{out}}e^{-\i\varphi}+\dd_{\text{out}}e^{\i\varphi}\right)/\sqrt{2}\,.\label{eq:Uout_quadrature_squeezing_angle}
\end{equation}
The fluctuations in this quantity are quantified by the (measurable)
spectral density:
\begin{equation}
S_{U_{\varphi}}^{\text{out}}\left[\omega\right]=\left\langle \int\mathrm{d}\tau\, e^{\i\omega\tau}\left\langle \hat{U}_{\varphi}^{\text{out}}\left(t+\tau/2\right)\hat{U}_{\varphi}^{\text{out}}\left(t-\tau/2\right)\right\rangle \right\rangle _{t}\,,\label{eq:def_output_light_spectrum}
\end{equation}
where $\left\langle \cdot\right\rangle _{t}$ denotes a time average
over the centre-of-mass time $t$ (i.e., we are interested in the
stationary part of the noise).

If the output light is in a coherent state, $\d_{{\rm out}}$ will
be in its vacuum, and $S_{U_{\varphi}}^{{\rm out}}[\omega]=1/2\equiv S_{SN}^{{\rm out}}$
(i.e., the shot-noise value); with the optomechanical interaction,
we will obtain deviations from this result. We will focus on the output
quadrature exhibiting the minimum noise at a given frequency $\omega$,
obtained by choosing the optimal angle $\varphi[\omega${]} (the squeezing
angle). Defining the orthogonal quadratures $\hat{U}_{1}^{{\rm out}}=\hat{U}_{\varphi=0}^{{\rm out}}$
and $\hat{U}_{2}^{{\rm out}}=\hat{U}_{\varphi=\pi/2}^{{\rm out}}$,
a straightforward optimization yields that the noise of this optimal
quadrature is (see, e.g., \cite{2012_Hammerer_NonclassicalStatesOfLightAndMechanics})
\begin{equation}
S_{\text{opt}}^{\text{out}}=\frac{2S_{U_{1}}^{\text{out}}S_{U_{2}}^{\text{out}}-2\left[S_{U_{1}U_{2}}^{\text{out}}\right]^{2}}{S_{U_{1}}^{\text{out}}+S_{U_{2}}^{\text{out}}+\sqrt{\left[S_{U_{1}}^{\text{out}}-S_{U_{2}}^{\text{out}}\right]^{2}+4\left[S_{U_{1}U_{2}}^{\text{out}}\right]^{2}}}\,.\label{eq:optimal_squeezing_ponderomotive}
\end{equation}
Here, the cross-correlator $S_{U_{1}U_{2}}^{\text{out}}\left[\omega\right]$
measures the classical (i.e., symmetrized) correlations between $\hat{U}_{1}^{{\rm out}}$
and $\hat{U}_{2}^{{\rm out}}$, and is defined as:
\begin{eqnarray*}
S_{U_{1}U_{2}}^{\text{out}}\left[\omega\right] & =\frac{1}{2} & \left\langle \int\mathrm{d}\tau\, e^{\i\omega\tau}\left\langle \hat{U}_{1}^{{\rm out}}\left(t+\tau/2\right)\hat{U}_{2}^{{\rm out}}(t-\tau/2)\right.\right.\\
 &  & \left.\left.+\hat{U}_{2}^{{\rm out}}(t+\tau/2)\hat{U}_{1}^{{\rm out}}(t-\tau/2)\right\rangle \right\rangle _{t}
\end{eqnarray*}

\section{Ponderomotive squeezing}

\textit{\emph{We begin by quickly reviewing the standard mechanism
for optomechanical squeezed light generation, ponderomotive squeezing
}}\cite{1967_Braginsky_PonderomotiveEffects,1994_Fabre_QuNoise,1997_Mancini_PonderomotiveControl,2012_Brooks_Non-classicalLight,2013_Safavi_Naeini_PonderomotiveSqueezing,2013_Purdy_PonderomotiveSqueezing}\textit{\emph{,
where one uses the}}\textit{ coherent} (i.e.,\textit{\emph{~Hamiltonian)
optical nonlinearity induced by the coupling to the mechanical resonator}}.
We assume a resonantly driven optomechanical cavity, i.e., $\alpha\left(t\right)=\alpha_{L}e^{-\i\omega_{{\rm cav}}t}$,
where $\alpha_{L}$ is the laser amplitude. Going into an interaction
picture with respect to the free cavity Hamiltonian and performing
a standard linearization on (\ref{eq:OM_Hamiltonian}) (i.e., dropping
terms cubic in $\hat{d},\hat{d}^{\dagger}$) one finds
\begin{equation}
\hat{H}=\hbar\Omega\bd\b-\sqrt{2}\hbar G\hat{U}_{1}\left(\hat{b}+\hat{b}^{\dagger}\right)\,.\label{eq:HResonantDrive}
\end{equation}
where $G=g_{0}\bar{a}$ is the drive-enhanced optomechanical coupling
strength; without loss of generality, we take the average cavity amplitude\textbf{
}$\bar{a}$ to be real. With this choice, $\hat{U}_{1}$ and $\hat{U}_{2}$
correspond respectively to standard amplitude and phase quadratures.
Their fluctuations are given by \cite{2012_Hammerer_NonclassicalStatesOfLightAndMechanics}
\begin{equation}
S_{U_{1}}^{\text{out}}=S_{\text{SN}}^{\text{out}}\,\text{ and}\, S_{U_{2}}^{\text{out}}=S_{U_{1}}^{\text{out}}+2\left[S_{U_{1}U_{2}}^{\text{out}}\right]^{2}+\delta S\,,\label{eq:SU1_SU2}
\end{equation}
where
\begin{eqnarray*}
\delta S & = & \tilde{S}(\tilde{S}+\coth\hbar\omega/2k_{B}T)\,,\\
\tilde{S} & = & 2\Omega G^{2}\kappa\mathrm{Im}\left\{ \chi_{M}\right\} /(\kappa^{2}/4+\omega^{2})\,,
\end{eqnarray*}
and
\[
\chi_{M}^{-1}=\Omega^{2}-\omega^{2}-\i\omega\Gamma_{M}
\]
is the mechanical susceptibility.

Given that neither $U_{1}$ nor $U_{2}$ is squeezed, obtaining squeezing
will necessarily require non-zero classical correlations between the
amplitude and phase quadrature (i.e., $S_{U_{1}U_{2}}^{\text{out}}\not=0$),
as follows from Eqs.~(\ref{eq:optimal_squeezing_ponderomotive})
and (\ref{eq:SU1_SU2}). These correlations are created by the mechanical
motion. From the last term of Eq.~(\ref{eq:HResonantDrive}), we
see that the amplitude ($U_{1}$) fluctuations of the light are a
driving force on the mechanics. The same term tells us that the resulting
mechanical motion modulates the phase of the light leaving the cavity
(i.e., the $U_{2}$ quadrature). One finds that the amplitude-phase
correlator has a simple form which completely reflects this intuitive
picture:
\begin{equation}
S_{U_{1}U_{2}}^{\text{out}}\left[\omega\right]\propto\frac{4G^{2}}{\kappa}\frac{\Omega}{1+\left(2\omega/\kappa\right)^{2}}\mathrm{Re}\left\{ \chi_{M}\left[\omega\right]\right\} \,,\label{eq:S_U1U2}
\end{equation}
where $\omega$ is measured in our rotating frame (i.e., $\omega=0$
corresponds to the cavity resonance). Note that only the real part
of $\chi_{M}$ enters, as only in-phase correlations between $U_{1}$
and $U_{2}$ are relevant to squeezing (i.e., the correlations are
induced by a \textit{coherent} interaction only, since the dissipative
part $\mathrm{Im}\left\{ \chi_{M}\right\} $ of $\chi_{M}$ does \textit{not}
enter). Such in-phase correlations between amplitude and phase quadratures
would naturally be created if we had a Kerr nonlinearity in the cavity,
i.e., a term $\hat{a}^{\dagger}\hat{a}^{\dagger}\hat{a}\hat{a}$ in
the cavity Hamiltonian. Thus, PS involves the optomechanical interaction
mimicking the effects of a (instantaneous, coherent, Hamiltonian)
Kerr interaction in the cavity. Note that the optomechanical interaction
was recently compared to a Kerr nonlinearity also in Refs.~\cite{2012_Kronwald_FCS,2013_Aldanda_Comparison_OM_system_Kerr_medium}.

\begin{figure}
\begin{centering}
\includegraphics[scale=1.25]{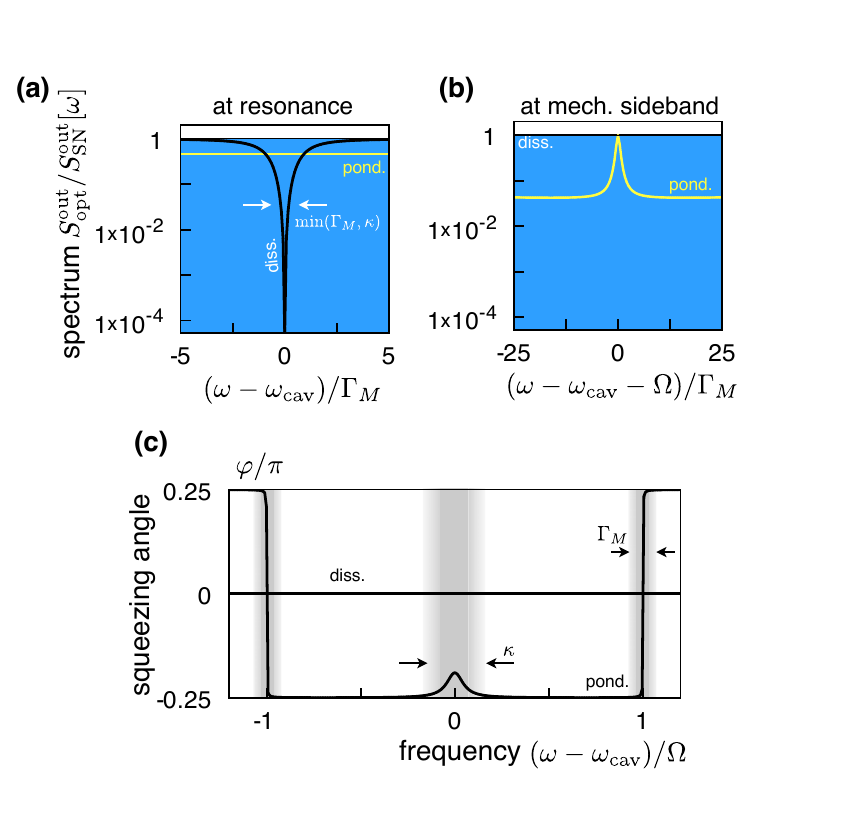}
\par\end{centering}

\caption{Ponderomotive (PS) vs.~dissipative squeezing spectra in the good
cavity limit (where the impedance matching condition $\tilde{\kappa}=\kappa$
is assumed, cf.~main text). (a) Output light spectra vs.~frequency
near the cavity resonance frequency $\omega_{\text{cav}}$ in the
good cavity limit (where $\kappa/\Omega\to0$ for our dissipative
scheme and $\kappa/\Omega=1/10$ for PS). The squeezing bandwidth
at $\omega_{\text{cav}}$ is set by $\kappa$ for PS. For our dissipative
scheme it is set by $\text{min}\left\{ \kappa,\Gamma_{M}\right\} $.
Both schemes generate maximum squeezing at $\omega_{\text{cav}}$
within this bandwidth. However, our dissipative scheme outperforms
PS in the good cavity limit. (b) Output light spectra at the mechanical
sideband $\omega\approx\Omega$. Our dissipative scheme does not generate
squeezing while PS does generates squeezing. (c) Squeezing angle $\varphi$
(cf.~Eq.~(\ref{eq:Uout_quadrature_squeezing_angle})) vs. frequency.
For dissipative squeezing, the squeezing angle is constant for all
frequencies. In contrast, the squeezing angle $\varphi_{\text{opt}}$
corresponding to optimal PS varies on a scale $\sim\kappa$ at the
cavity resonance and $\sim\Gamma_{M}$ close to the mechanical sideband.
{[}Parameters: (a) $\Gamma_{M}=2\cdot10^{-5}\kappa,\nth=10$, and
$\C=10^{5}$ (b) Same as (a), $\tilde{\kappa}=4\G^{2}/\Gamma_{M}=\kappa$.(c)
Same as in (a){]}}

\label{FIG_2}
\end{figure}

It thus follows that PS will be strongest at frequencies $\omega$,
where the correlator $S_{U_{1}U_{2}}^{\text{out}}$ is large; by combining
Eqs.~(\ref{eq:optimal_squeezing_ponderomotive}) and (\ref{eq:SU1_SU2}),
one finds $S_{\text{opt}}^{\text{out}}\propto1/[S_{U_{1}U_{2}}^{\text{out}}]^{2}$
for $S_{U_{1}U_{2}}^{\text{out}}\gg1$. The correlations will in turn
be large when the real part of the mechanical susceptibility is large.
This naturally occurs at the cavity resonance frequency (i.e., $\omega=0$
in Eq.~(\ref{eq:S_U1U2})), and also near (but not at) the mechanical
sideband frequencies, i.e.,~frequencies $\omega=\pm\Omega+\delta$
where $\left|\delta\right|\sim\Gamma_{M}$. Fig.~\ref{FIG_2} shows
this expected frequency dependence of ponderomotive squeezing. 

It is often overlooked that the same intuition used above tells us
that PS will be suppressed in the good-cavity limit $\kappa\ll\Omega$
, a limit necessary for ground-state optomechanical cooling and other
desirable optomechanical protocols. At the cavity resonance, $S_{U_{1}U_{2}}^{\text{out}}\propto4G^{2}/\left(\kappa\Omega\right)$,
independent of the sideband parameter $\kappa/\Omega$ and mechanical
damping rate $\Gamma_{M}$. Thus, in the limit $\Omega/\kappa\to\infty$
while $G/\kappa$ remains fixed, ponderomotive squeezing disappears
at the cavity frequency. Indeed, we find $S_{\text{opt}}^{\text{out}}/S_{\text{SN}}^{\text{out}}\approx1-16G^{2}/\left(\kappa\Omega\right)$
in this limit. The situation is different for frequencies close to
the mechanical sidebands, i.e., $\omega=\pm\Omega+\delta$ with $\left|\delta\right|\sim\Gamma_{M}$.
In the bad cavity limit $\kappa\gg\Omega$, $S_{U_{1}U_{2}}^{\text{out}}\propto\C$,
where the cooperativity $\C=4G^{2}/\left(\kappa\Gamma_{M}\right)$.
Thus, squeezing close to the mechanical sideband and for $\kappa\gg\Omega$
is controlled by the cooperativity only. In the good cavity limit
$\kappa\ll\Omega$, however, $S_{U_{1}U_{2}}^{\text{out}}\propto\C\left(\kappa/\Omega\right)^{2}$.
Thus, in the good cavity limit $\kappa/\Omega\to0$, squeezed light
cannot be generated effectively using the standard ponderomotive squeezing
mechanism.

\section{Dissipative output light squeezing}

Given the general desirability of having optomechanical systems in
the good-cavity limit (e.g.~for cooling \cite{2007_Wilson-Rae_TheoryGroundStateCooling,2007_FM_SidebandCooling,2008_Schliesser_SidebandCooling,2011_Teufel_SidebandCooling_Nature,2011_Chan_LaserCoolingNanomechOscillator},
state transfer \cite{2011_Fiore_QuantumStateTransfer,2011_Safavi-Naeini_PPT,2012_Wang_QuantumStateTransfer,2012_Verhagen_QuantumCoherentCoupling,2013_Palomaki_CohStateTransfer},
entanglement generation \cite{2013_Palomaki_EntanglingMechMotionWMicrowaveFields,2007_Vitali_OMEntanglement,2007_Paternostro_EntanglementWithLight,2012_Schmidt_CVQuantumStateProcessing,2013_WangClerk_SteadyStateEntanglement},
etc.), it would be extremely useful to find an alternative squeezing
scheme which is efficient in this regime. To that end, we now introduce
an approach which generates squeezed light by explicitly using the
dissipative nature of the mechanical resonator.

\subsection{Basic scheme}

Unlike PS, the dissipative approach to optomechanical squeezing requires
driving the cavity with two lasers, with frequencies corresponding
to the red and blue mechanical sidebands (i.e.,~$\alpha\left(t\right)=\alpha_{+}e^{-\i\left(\omega_{\text{cav}}+\Omega\right)t}+\alpha_{-}e^{-\i\left(\omega_{\text{cav}}-\Omega\right)t}$);
the resulting average classical amplitude is $\bar{a}(t)=e^{-i\omega_{{\rm cav}}t}\sum_{\sigma=\pm}\bar{a}_{\sigma}e^{-i\sigma\Omega t}$.
We again write the basic optomechanical Hamiltonian of Eq.~(\ref{eq:OM_Hamiltonian})
in an interaction picture, now with respect to both the free cavity
and mechanical resonator Hamiltonians. Introducing mechanical quadrature
operators $\hat{X}_{1}=\left(\bd+\b\right)/\sqrt{2}$ and $\hat{X}_{2}=\i\left(\bd-\b\right)/\sqrt{2}$,
and linearizing the Hamiltonian in the usual way, we find $\hat{H}=\hat{H}_{S}+\hat{H}_{CR}$,
where
\begin{eqnarray}
\hat{H}_{S} & = & -\hbar\left(G_{+}+G_{-}\right)\hat{U}_{1}\hat{X}_{1}-\hbar\left(G_{-}-G_{+}\right)\hat{U}_{2}\hat{X}_{2},\label{eq:Hsqueezing}\\
\hat{H}_{\text{CR}} & = & -\hbar\dd\left(G_{+}\b e^{-2\i\Omega t}+G_{-}\bd e^{2\i\Omega t}\right)+\text{h.c.}.\label{eq:H_CR_1}
\end{eqnarray}
Here $G_{\pm}=g_{0}\bar{a}_{\pm}$ are the many-photon optomechanical
couplings associated with each drive tone; we take $\bar{a}_{+},\bar{a}_{-}$
to be real and positive without any loss of generality. The terms
in $\hat{H}_{S}$ describe resonant interaction processes that will
give rise to squeezing, while those in $\hat{H}_{\text{CR}}$ are
deleterious non-resonant interaction terms. For physical transparency,
we will start by discussing the extreme good cavity limit $\kappa\ll\Omega$,
and thus ignore the effects of $\hat{H}_{\text{CR}}$. We will also
take $G_{-}\geq G_{+}$, which ensures the stability of the linearized
system. 

If $G_{+}=G_{-}$, $\hat{H}_{S}$ has the form of a quantum non-demolition
(QND) interaction, as both the quadratures $U_{1}$ and $X_{1}$ commute
with the Hamiltonian; such a regime can be used to make a back-action
evading measurement of the mechanical quadrature $X_{1}$ \cite{1980_Caves_ForceMsmntQuantumMechOsc,1980_Braginsky_QND_Quadrature_Measurement,2008_Clerk_BackActionEvasion}.
For $G_{+}\neq G_{-}$, the second term in $\hat{H}_{S}$ is non-zero,
and the QND structure is lost. As we recently discussed \cite{2013_Kronwald_MechanicalSqueezing},
this regime can be extremely efficient for the generation of mechanical
squeezing. 

Given that Eq.~(\ref{eq:Hsqueezing}) is symmetric under interchange
of mechanical and cavity quadratures, one might naturally suspect
that it can also be exploited to generate optical squeezing. We now
show that this is indeed the case, even though in the optical case,
we are interested in squeezing a quadrature of the \emph{output} light
field, not the intracavity field. As is well known, the relationship
between intracavity and output field squeezing can be non-trivial
\cite{1984_Collett_Squeezing,1990_GeaBanacloche_OutputLightSqueezing_Not_IntracavityLightSqueezing}.
We show that this is also the case here.

\subsection{Underlying mechanism}

We start by describing the basic mechanism which gives rise to squeezing
here, considering the most interesting regime where $0<G_{-}-G_{+}\ll G_{-}+G_{+}$;
for simplicity, we also first consider the case of a large mechanical
damping rate $\Gamma_{M}\gg\kappa$. The first term in $\hat{H}_{S}$
(cf.~Eq.~(\ref{eq:Hsqueezing})) causes the mechanical resonator's
$X_{2}$ quadrature to measure the cavity $U_{1}$ quadrature: In
the relevant low-frequency limit, one finds 
\[
\hat{X}_{2}=2\frac{G_{+}+G_{-}}{\Gamma_{M}}\hat{U}_{1}+\frac{2}{\sqrt{\Gamma_{M}}}\hat{X}_{2}^{\text{in}}\,.
\]
Thus, the measurement strength $\propto G_{-}+G_{+}$. This also demonstrates
that dissipation is a necessary ingredient for $\hat{X}_{2}$ to measure
the $\hat{U}_{1}$ quadrature. In contrast, the second term in $\hat{H}_{S}$
perturbs the measured quadrature $U_{1}$ by applying a weak force
$\propto\left(G_{-}-G_{+}\right)\hat{X}_{2}$. However, as $X_{2}$
has measured $U_{1}$, this becomes a weak feedback force.

The result of these two operations is a net additional damping of
the cavity $U_{1}$ quadrature at rate $\tilde{\kappa}=4\G^{2}/\Gamma_{M}$
due to the optomechanical interaction, where $\G^{2}=G_{-}^{2}-G_{+}^{2}$.
The mechanical resonator is thus acting like a dissipative bath for
the cavity photons. One must also ask about the extra noise introduced
into the cavity quadrature $U_{1}$ via the optomechanical coupling.
As this only involves the weak second term in $\hat{H}_{S}$ ($\propto(G_{-}-G_{+})$,
cf.~Eq.~(\ref{eq:Hsqueezing})), this noise is extremely small,
much smaller than the noise we would expect if $\tilde{\kappa}$ was
produced by a zero-temperature dissipative bath. The net result is
that the mechanical resonator acts as a squeezed bath for the cavity,
damping the $U_{1}$ quadrature while adding almost no fluctuations.
This directly causes optical squeezing. The situation is of course
reversed if we now ask about the cavity $U_{2}$ quadrature. As the
measurement and feedback roles of the two terms in $\hat{H}_{S}$
are reversed for $U_{2}$, its fluctuations are naturally enhanced
by the effective mechanical bath.

\subsection{Detailed calculation\label{sub:Detailed-calculation}}

The above picture provides intuition for how the combination of the
Hamiltonian $\hat{H}_{s}$ in Eq.~(\ref{eq:Hsqueezing}) and mechanical
damping gives rise to squeezing of the intracavity field: the mechanical
resonator (via autonomous measurement and feedback operations) mimics
the actions of squeezed dissipative reservoir coupled to the cavity.
To understand how this basic mechanism affects the output noise of
the cavity, we simply solve the linearized equations of motion describing
our system (now without any assumption of a large $\Gamma_{M}$). 

To present the solutions in a transparent manner, we first introduce
the self-energy of the cavity photons due to the optomechanical interaction
and the corresponding dressed cavity susceptibility:
\begin{eqnarray*}
\Sigma\left[\omega\right] & = & \frac{-\i\left(G_{-}^{2}-G_{+}^{2}\right)}{-\i\omega+\Gamma_{M}/2}\equiv\textrm{Re}\,\,\Sigma[\omega]-i\tilde{\kappa}[\omega]/2.
\end{eqnarray*}
The corresponding dressed cavity susceptibility (Green function) is
then

\begin{eqnarray}
\chi_{{\rm cav}}[\omega] & = & \frac{1}{-i\omega+(\kappa/2)+i\Sigma[\omega]}.\label{eq:dressed_cav_suscept}
\end{eqnarray}
The output cavity quadrature operators are then found to be
\begin{eqnarray}
\hat{U}_{1}^{\text{out}}\left[\omega\right] & = & \left(\kappa\chi_{{\rm cav}}[\omega]-1\right)\hat{U}_{1}^{\text{in}}\left[\omega\right]-\sqrt{\kappa\Gamma_{M}}\chi_{{\rm cav}}[\omega]\sqrt{\tilde{\kappa}[\omega]}\hat{\xi}_{1}\left[\omega\right],\label{eq:U1out}\\
\hat{U}_{2}^{\text{out}}\left[\omega\right] & = & \left(\kappa\chi_{{\rm cav}}[\omega]-1\right)\hat{U}_{2}^{\text{in}}\left[\omega\right]+\sqrt{\kappa\Gamma_{M}}\chi_{{\rm cav}}[\omega]\sqrt{\tilde{\kappa}[\omega]}\hat{\xi}_{2}\left[\omega\right].\label{eq:U2out}
\end{eqnarray}

These input/output relations have the expected simple form for a cavity
which is coupled both to a coupling port (coupling rate $\kappa$)
and to an additional dissipative reservoir (coupling rate $\tilde{\kappa}[\omega]$).
The coupling to the additional reservoir both modifies the cavity
susceptibility, and results in new driving noises. The first term
on the RHS of Eqs.~(\ref{eq:U1out}-\ref{eq:U2out}) corresponds
to the contribution to the output field from vacuum noise incident
on the cavity from the coupling port: there is both a promptly reflected
contribution, and a contribution where this noise enters the cavity
before being emitted. Note that these terms are completely phase insensitive,
i.e.,~identical in form for any choice of optical quadrature. 

More interesting are the second terms on the RHS of Eqs.~(\ref{eq:U1out}-\ref{eq:U2out}),
which represent the noise contributions from the effective mechanical
bath coupled to the cavity. One finds
\begin{eqnarray*}
\hat{\xi}_{1/2} & = & \frac{1}{\sqrt{\tilde{\kappa}[\omega}]}\frac{G_{-}\mp G_{+}}{-i\omega+\Gamma_{M}/2}\hat{X}_{2/1}^{{\rm in}}[\omega]
\end{eqnarray*}
We see immediately that this effective bath seen by the cavity appears
squeezed (i.e.,~the noise in $\hat{\xi}_{1}$ is much less than that
in $\hat{\xi}_{2}$) even if the intrinsic mechanical dissipation
is in a simple thermal state. 

With these equations, the route towards optimal squeezing at frequency
$\omega$ is clear: one needs both to have $G_{-}-G_{+}$ be as small
as possible (so that the $\hat{\xi}_{j}$ noises are as squeezed as
possible), while at the same time fulfilling an impedance matching
condition that makes the first terms in Eqs.~(\ref{eq:U1out}-\ref{eq:U2out})
vanish, i.e., $\kappa\chi_{{\rm cav}}[\omega]=1$. Physically, this
impedance matching simply means that all the incident optical vacuum
fluctuations on the cavity are completely absorbed by the mechanical
resonator, cf.~Fig.~\ref{FIG_1}b. At the cavity resonance frequency
($\omega=0),$this corresponds to a simple matching of damping rates
\begin{eqnarray}
\tilde{\kappa}[0]=\kappa & \Longleftrightarrow & \frac{4\left(G_{-}^{2}-G_{+}^{2}\right)}{\Gamma_{M}}=\kappa\label{eq:impedance_matching-1}
\end{eqnarray}
We also see that regardless of the frequency we consider, the $U_{1}[\omega${]}
optical quadrature is the optimally squeezed quadrature; this is simply
because the squeezing angle of our effective mechanical bath is frequency
independent.

\subsection{\textit{\emph{Results}}}

Having explained the basic dissipative squeezing mechanism, we now
present results for the amount of generated squeezing, again starting
with the extreme good cavity limit $\Omega\gg\kappa$. The simplest
regime here is the weak-coupling regime, where the effective coupling
$\mathcal{G}=\sqrt{G_{-}^{2}-G_{+}^{2}}$ is much smaller than $\max(\Gamma_{M},\kappa)$.
The output light is maximally squeezed at the cavity frequency, cf.~Fig.~\ref{FIG_2};
the squeezing remains appreciable away from the cavity resonance over
a {}``squeezing bandwidth'' set by $\mathrm{max}\left\{ \kappa,\Gamma_{M}\right\} $.
 The amount of squeezing at the cavity resonance is given by
\begin{equation}
\frac{S_{U_{1}}^{\text{out}}\left[\omega=0\right]}{S_{\text{SN}}^{\text{out}}}=\frac{4\kappa\tilde{\kappa}\left(1+2\nth\right)e^{-2r}+\left(\kappa-\tilde{\kappa}\right)^{2}}{\left(\kappa+\tilde{\kappa}\right)^{2}}\,,\label{eq:squeezing_omega_0}
\end{equation}
where we have introduced the squeezing parameter $r$ via $\tanh r=G_{+}/G_{-}$,
i.e., the ratio of laser drive amplitudes. Note that this expression
is valid in the extreme good cavity limit $\kappa/\Omega\to0$ for
all values of $\kappa,\tilde{\kappa}$ and $r$\textit{.} For a fixed
squeezing parameter $r$, the noise in the $U_{1}$ quadrature interpolates
between three simple limits. For $\tilde{\kappa}=0$ or $\tilde{\kappa}\gg\kappa$,
the noise of the effective mechanical resonator is completely reflected
from the cavity, and hence the output quadrature noise is the just
vacuum noise of the incident field. In contrast, if the impedance
matching condition of Eq.~(\ref{eq:impedance_matching-1}) is satisfied,
then the output optical noise is completely determined by the effective
mechanical bath; it thus has the value $(1+2n_{{\rm th}})e^{-2r}$,
reflecting the effective temperature of the squeezed $\hat{\xi}_{1}$
noise associated with the effective mechanical bath. 

The above result then implies that for the optimal impedance-matched
case (which also implies being in the assumed weak-coupling regime,
cf.~appendix \ref{sub:SqueezingInTheStrongCouplingRegime}), the
squeezing of the cavity light at resonance behaves as 
\begin{equation}
S_{U_{1}}^{\text{out}}\left[0\right]/S_{\text{SN}}^{\text{out}}=\left(1+2\nth\right)e^{-2r}\approx\frac{1+2\nth}{4\C}\label{eq:squeezing_omega_0_imped_matched}
\end{equation}
where we have introduced the optomechanical cooperativity $\mathcal{C}=4G_{-}^{2}/\kappa\Gamma_{M}$
, and in the last expression we assumed $\C\gg1$. 

It is also interesting to consider the purity of the output light
generated; not surprisingly, for the optimal impedance matched case,
this purity is completely determined by the purity of the mechanical
noise. Parameterizing the purity of the output light via an effective
number of thermal quanta $n_{{\rm eff}}$, i.e., 
\[
\left(1+2\neff\left[\omega\right]\right)^{2}=4S_{U_{1}^{\text{out}}U_{1}^{\text{out}}}\left[\omega\right]S_{U_{2}^{\text{out}}U_{2}^{\text{out}}}\left[\omega\right]\,,
\]
one finds $\neff=\nth$ at the cavity frequency $\omega=0$ and for
$\tilde{\kappa}=\kappa$.

\subsection{Dissipative vs. ponderomotive squeezing}

Let us now compare our dissipative scheme to ponderomotive squeezing
(PS). PS squeezes light by correlating the incident optical vacuum
fluctuations using the \textit{coherent} Kerr interaction mediated
by the mechanical resonator. In contrast, our approach does not rely
on correlating the incident optical vacuum fluctuation; rather, we
replace these fluctuations by squeezed noise emanating from the mechanical
resonator. As discussed, our scheme also relies crucially on the \textit{dissipative}
nature of the mechanical resonator, i.e.,~on the imaginary part of
the mechanical susceptibility $\chi_{M}$. In contrast, a non-vanishing
$\mathrm{Im}\chi_{M}$ reduces the amount of ponderomotive squeezing,
cf.~Eq.~(\ref{eq:SU1_SU2}). We also note that our scheme is efficient
in the good cavity limit and generates squeezing with a fixed squeezing
angle, in contrast to PS.

\begin{figure}
\begin{centering}
\includegraphics[scale=1.25]{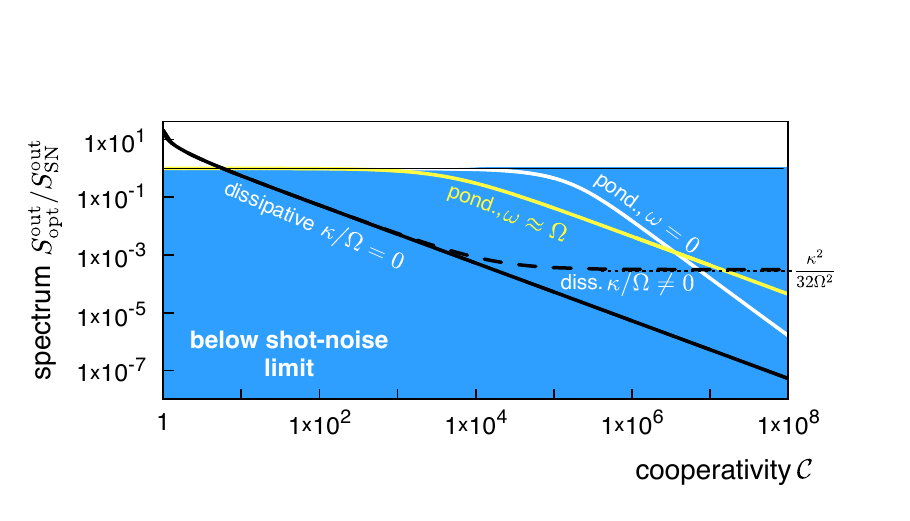}
\par\end{centering}

\caption{Dissipative vs. ponderomotive squeezing in the good cavity limit.
Black curves: Output spectrum for the dissipative squeezing scheme
at $\omega=0$, i.e., at the cavity resonance frequency (black solid
curve for $\kappa/\Omega=0$ and black dashed curve for $\kappa/\Omega=1/10$$ $).
White, yellow curves: optimized spectrum for standard ponderomotive
squeezing, for both a frequency $\omega=0$ and a frequency $\omega\sim\Omega$.
Note that the value of the output spectrum for dissipative squeezing
for small cooperativities is set by $n_{\text{th}}$ {[}Parameters
as in Fig. \ref{FIG_2}(a,b){]}.}

\label{FIG_3}
\end{figure}

Let us now turn to a \textit{quantitative }comparison of our dissipative
scheme to ponderomotive squeezing in the good cavity limit $\kappa\ll\Omega$,
cf.~Fig.~\ref{FIG_3}. We parametrize the red laser strength (or
the resonant laser strength for PS) via the cooperativity $\C=4G_{-}^{2}/\left(\kappa\Gamma_{M}\right)$
(where $G_{-}\mapsto G$ for ponderomotive squeezing). For our dissipative
scheme, we optimize the blue laser strength for any given cooperativity
to fulfill the impedance matching condition (\ref{eq:impedance_matching-1}). 

We now compare the amount of squeezing generated by our dissipative
scheme at the cavity frequency, i.e., $S_{U_{1}}^{\text{out}}\left[0\right]$
to PS, i.e., to the optimized output light spectrum $S_{\text{opt}}^{\text{out}}$
at the cavity frequency and close to the mechanical sideband. For
small cooperativities, $1<\C<\left(1+\nth\right)^{2}/\left(1+2\nth\right)$,
the output light spectrum in our scheme corresponds to thermally squeezed
light (as $S_{U_{1}}^{\text{out}}\left[0\right]/S_{\text{SN}}^{\text{out}}<(1+2\nth)$).
As the squeeze parameter is small in this regime (cf.~Eq.~(\ref{eq:squeezing_omega_0_imped_matched})),
$S_{U_{1}}^{\text{out}}$ is larger than the shot noise value. In
contrast, the output light spectrum $S_{\text{opt}}^{\text{out}}$
for PS in this small-$\C$ case stays close to the shot-noise limit
as $S_{\text{opt}}^{\text{out}}\approx S_{\text{SN}}^{\text{out}}$.
As soon as $\C\gtrsim\nth/2$, our scheme generates \textit{quantum
}squeezed output light where $S_{U_{1}}^{\text{out}}\left[0\right]<S_{\text{SN}}^{\text{out}}$.
PS, however, still stays close to the shot-noise limit, $S_{\text{opt}}^{\text{out}}\approx S_{\text{SN}}^{\text{out}}$.
While increasing the cooperativity further, PS also starts to generate
strong quantum squeezing, first close to the mechanical sideband,
then also at $\omega=0$. Thus, when comparing our scheme to ponderomotive
squeezing for a fixed cooperativity, we see that our scheme outperforms
ponderomotive squeezing in the good cavity limit. This can also be
seen by studying the minimum cooperativity $\C_{\text{min}}$ needed
to generate a certain amount of squeezing, e.g. $3\,\mathrm{dB}$.
For our scheme, we find $\C_{\text{min}}^{\text{diss}}\gtrsim\left(1+2\nth\right)/2$.
In contrast, for ponderomotive squeezing in the good cavity limit
we find $\C_{\text{min}}^{\text{PS}}\gtrsim\left(\C_{\text{min}}^{\text{diss}}+\Omega/\left(\sqrt{2}\Gamma_{M}\right)\right)/4$.
This is typically much larger than $\C_{\text{diss}}$ since $\Gamma_{M}\ll\Omega$
for typical experiments.

\subsection{Bad cavity effects on the generation of squeezed output light\label{sub:Bad-cavity-effects-squeezed-output-light}}

Up to now, we have focussed on the extreme good cavity limit, i.e.,
$\kappa/\Omega\to0$. We now consider deviations that arise when $\kappa/\Omega$
is non-zero. 

Thus, we now solve the full quantum Langevin equations including $\hat{H}_{\text{CR}}$
(i.e., no rotating-wave approximation) and analyze the output light
spectrum $S_{U_{1}}^{\text{out}}\left[\omega\right]$. We find that
the impedance matching condition $\tilde{\kappa}=\kappa$ still maximizes
squeezing at the cavity frequency $\omega=0$. Thus, we now compare
$S_{U_{1}}^{\text{out}}\left[0\right]$ with and without bad cavity
effects, cf.~Fig.~\ref{FIG_3}. The amount of squeezing for moderate
cooperativities does not differ from the good cavity prediction (\ref{eq:squeezing_omega_0_imped_matched}).
As the cooperativity gets larger, however, the impact of bad cavity
effects also becomes larger. As these terms tend to heat up the cavity
quadrature non-resonantly, the maximum amount of squeezing our dissipative
scheme can generate is limited. By taking $\hat{H}_{\text{CR}}$ into
account up to leading order in $\kappa/\Omega$, we find that in the
large cooperativity limit 
\[
\frac{S_{U_{1}}^{\text{out}}\left[0\right]}{S_{\text{SN}}^{\text{out}}}=\frac{\kappa^{2}}{32\Omega^{2}}\,,
\]
where $G_{+}/G_{-}$ was again chosen to fulfill the impedance matching
condition (\ref{eq:impedance_matching-1}).

\section{Increasing the measurement sensitivity of an optomechanical cavity\label{sec:Increasing-the-measurement}}

As we have seen, our dissipative scheme can be used to generate squeezed
output light. This light could then be fed into a separate measurement
device to increase its measurement sensitivity. Such a scenario, however,
involves two different devices which have to be coupled. In order
to avoid unwanted coupling losses which could degrade the measurement
sensitivity again or to keep the experiment as simple as possible,
one might ask whether the squeezed light source and the measurement
device could somehow be combined. In the following, we show that this
is indeed possible: one could use the optomechanical cavity to both
generate squeezed output light while increasing the sensitivity for
measuring a dispersively-coupled signal at the same time.

\subsection{Basic scheme}

We now consider an optomechanical cavity which is also dispersively
coupled to a signal $z$ we want to measure (one could e.g.~use an
optomechanical setup in the microwave regime where a superconducting
qubit is dispersively coupled to the microwave cavity; $z$ would
then be a Pauli operator $\sigma_{z}$ for the qubit). We again assume
two lasers driving the cavity on the red and blue mechanical sideband.
As discussed above, the corresponding optomechanical interaction will
cause the $\hat{U}_{1}^{\text{out}}$-quadrature to be squeezed at
the cavity frequency. We now also add a resonant measurement tone
which is used to probe the value of $z$. Thus,

\[
\hat{H}=\hbar\omega_{\text{cav}}\ad\a+\hbar\Omega\bd\b-\hbar g_{0}\left(\bd+\b\right)\ad\a-\hbar A\ad\a\cdot z+\hat{H}_{\text{dr}}\,,
\]
where $\hat{H}_{\text{dr}}=\hbar(\alpha\left(t\right)\ad+\text{h.c.})$
and $\alpha\left(t\right)=\alpha_{+}e^{-\i(\omega_{\text{cav}}+\Omega)t}+\alpha_{-}e^{-\i(\omega_{\text{cav}}-\Omega)t}+\alpha_{0}e^{-\i\omega_{\text{cav}}t}$.
Note that the measurement tone at frequency $\omega_{\text{cav}}$
is spectrally very well resolved from the two tones at frequency $\omega_{\pm}=\omega_{\text{cav}}\pm\Omega$
used to generate squeezing. Thus, we expect the measurement tone to
probe $z$ only without strongly degrading squeezing. 

We now apply the displacement transformation $\a=\bar{a}(t)+\d$ with
$\bar{a}(t)=e^{-\i\omega_{\text{cav}}t}(\sum_{\sigma=\pm}\bar{a}_{\sigma}e^{-i\sigma\Omega t}+\i\bar{a}_{0})$.
We also assume $\bar{a}_{i}$ to be real. Note that the phase of the
measurement tone is chosen such that the information of $z$ is imprinted
in the squeezed quadrature, as we will see below. This is crucial
to enhance the measurement sensitivity of the optomechanical cavity.
We go into a rotating frame with respect to the free cavity and mechanical
resonator Hamiltonian and apply standard linearization. We find $\hat{H}=\hat{H}_{S}+\hat{\tilde{H}}_{\text{CR}}$
with
\begin{equation}
\hat{H}_{S}=-\hbar\left(G_{+}+G_{-}\right)\hat{U}_{1}\hat{X}_{1}-\hbar\left(G_{-}-G_{+}\right)\hat{U}_{2}\hat{X}_{2}-\hbar\sqrt{2}A_{0}\hat{U}_{2}\cdot z\,,\label{eq:H_squeezing-1}
\end{equation}

and
\begin{eqnarray}
\hat{\tilde{H}}_{\text{CR}} & = & \hat{H}_{\text{CR}}-2\hbar G_{0}\left(\Xone\cos\Omega t+\Xtwo\sin\Omega t\right)\hat{U}_{2}\label{eq:H_CR_incl_measurement_tone}\\
 & -\sqrt{2}\hbar z & \left[\left(A_{+}+A_{-}\right)\hat{U}_{1}\cos\Omega t+\left(A_{-}-A_{+}\right)\hat{U}_{2}\sin\Omega t\right]\nonumber 
\end{eqnarray}
where $\hat{H}_{\text{CR}}$ is given by Eq.~(\ref{eq:H_CR_1}).
Here, $G_{i}=g_{0}\bar{a}_{i}$ is the driven-enhanced optomechanical
coupling whereas $A_{i}=A\bar{a}_{i}$ is the driven-enhanced dispersive
cavity-signal coupling. As in Sec. \ref{sub:Bad-cavity-effects-squeezed-output-light},
$\hat{\tilde{H}}_{\text{CR}}$ represents non-resonant interaction
terms that will have minimal effect in the $\kappa/\Omega\to0$ limit.

\subsection{Enhanced measurement rate}

Let us first focus on the extreme good cavity limit and ignore $\hat{\tilde{H}}_{\text{CR}}$. The
last term in Eq.~(\ref{eq:H_squeezing-1}) implies that the $\hat{U}_{1}$
cavity quadrature measures $z$. Thus, the value of $z$ can be inferred
by observing the output light quadrature $\langle\hat{I}\rangle=\sqrt{\kappa}\langle\hat{U}_{1}^{\text{out}}\rangle$
by using a homodyne measurement setup for instance.

As we are interested in a weak coupling between the cavity and the
signal $z$, it will take a finite amount of time $\tau_{\text{meas}}$
to resolve the value of $z$ above the noise. This measurement time
is quantified in the standard manner by the measurement time or rate
$\Gamma_{\text{meas}}=1/\tau_{\text{meas}}$ \cite{2008_ClerkDevoretGirvinFMSchoelkopf_RMP}.
The measurement rate is related to the (zero frequency) susceptibility
$\chi_{\text{meas}}\equiv\mathrm{d}\langle\hat{I}\rangle\left[\omega\right]/\mathrm{d}z$
and the symmetrized spectrum $\bar{S}_{II}$ of the homodyne current
$\hat{I}$ at zero frequency via

\begin{equation}
\Gamma_{\text{meas}}=\frac{\chi_{\text{meas}}^{2}}{2\bar{S}_{II}\left[0\right]}\,.\label{eq:def_msmnt_rate}
\end{equation}
Here, $\chi_{\text{meas}}[0]$ defines how much the average homodyne
current changes when $z$ is statically changed and the symmetrized
spectrum $\bar{S}_{II}$ quantifies the imprecision noise. 

We now see the route towards an enhanced measurement rate: we simply
need to use the optomechanical interaction and the consequent dissipative
squeezing mechanism to squeeze $\hat{U}_{1}^{\text{out}}$ at zero
frequency, and hence reduce $\bar{S}_{II}$ while keeping the measurement
susceptibility $\chi_{\text{meas}}$ as large as possible. 

As before, the optomechanical coupling in Eq.~(\ref{eq:H_squeezing-1})
generates squeezed output light where $\hat{U}_{1}^{\text{out}}$
is squeezed at $\omega=0$, i.e., at the cavity frequency. This directly
reduces the imprecision noise since $\bar{S}_{II}\left[\omega\right]=2\kappa S_{U_{1}}^{\text{out}}\left[\omega\right]$
such that $\bar{S}_{II}\left[0\right]=\kappa\left(1+2\nth\right)e^{-2r}$
for the impedance matching condition $\tilde{\kappa}=\kappa$, cf.~Eq.~(\ref{eq:squeezing_omega_0_imped_matched}).
At the same time, the measurement susceptibility $\chi_{\text{meas}}=-\sqrt{2}\kappa A_{0}\chi_{\text{cav}}\left[0\right]$
is not drastically changed. This is because when we optimally impedance
match to maximize squeezing, i.e., choosing $\tilde{\kappa}=\kappa$,
the optomechanical interaction only doubles the effective cavity damping,
cf.~Eq.~(\ref{eq:dressed_cav_suscept}). Thus, $\chi_{\text{cav}}\left[0\right]=1/\kappa$
is reduced only by a factor $1/2$ as compared to the value one would
obtain without the optomechanical interaction. Thus, we finally find
\[
\Gamma_{\text{meas}}=\frac{A_{0}^{2}}{\kappa}\frac{e^{2r}}{1+2\nth}\,.
\]

To quantify the sensitivity of our optomechanical cavity to $z$,
we compare this measurement rate to the rate $\Gamma_{\text{meas}}^{\text{lc}}$
we expect when $z$ is measured using a linear cavity. This corresponds
to turning off the optomechanical interactions in our scheme (i.e.,
$g_{0}\to0$). Hence, this comparison can be understood as being a
benchmark for our dissipative squeezing scheme. We find

\begin{eqnarray}
\frac{\Gamma_{\text{meas}}}{\Gamma_{\text{meas}}^{\text{lc }}} & = & \left(\frac{\chi_{\text{cav}}[0]}{\chi_{\text{cav}}^{\text{lc }}[0]}\right)^{2}\frac{1}{S_{U_{1}}^{\text{out,diss}}/S^{\text{SN}}}\label{eq:msmnt_rate_cf_linear_cavity}\\
 & = & \frac{e^{2r}}{4\left(1+2\nth\right)}\approx\frac{\C}{1+2\nth}\,,
\end{eqnarray}
where the last term is valid in the large $\C$ limit. Here, $\chi_{\text{cav}}$
is the dressed cavity susceptibility (cf.~Eq.~(\ref{eq:dressed_cav_suscept}))
and $\chi_{\text{cav}}^{\text{lc}}\left[0\right]=2/\kappa$ is the
susceptibility of a linear cavity at zero frequency. Thus, our scheme
allows for an exponential enhancement of the measurement rate with
the squeezing parameter $r$ (or a linear enhancement with cooperativity)
as long as $\C\gtrsim1+2\nth$. For this comparison we have assumed
equal decay rates $\kappa$ and the same read-out laser amplitudes. 

The above analysis demonstrates that our dissipative optomechanical
squeezing scheme can directly be used to enhance the intrinsic measurement
sensitivity. The crucial trick allowing this direct enhancement is
that our scheme generates squeezed output light without lowering the
(dressed) cavity susceptibility drastically. Additionally, the cavity
susceptibility is modified in a phase insensitive way, i.e., it is
identical for all quadratures, cf.~Eqs.~(\ref{eq:U1out},\ref{eq:U2out}). 

Note that it would be much more difficult to increase the intrinsic
measurement sensitivity using ponderomotive squeezing: There, the
optomechanical interaction effectively generates a Kerr-type optical
nonlinearity \cite{1994_Fabre_QuNoise,1997_Mancini_PonderomotiveControl}.
The corresponding linearized dynamics is similar to the dynamics of
a parametric amplifier. Squeezing is generated by modifying the cavity
susceptibility in a phase sensitive manner: one reduces the cavity
response to vacuum noise for one quadrature while increasing the response
for the conjugate quadrature. Reducing the response of the squeezed
quadrature to noise, however, will also reduce its response to the
signal $z$. Thus, the measurement rate $\Gamma_{\text{meas}}$  could
be unchanged.

\subsection{Influence of bad cavity effects on the measurement rate}

Let us now discuss the influence of bad cavity effects on the measurement
rate. Thus, we solve the quantum Langevin equations including $\hat{\tilde{H}}_{\text{CR}}$
numerically and analyze the corresponding output light spectrum. Note
first that the counter rotating terms which are independent of the
mechanical resonator (cf.~second line of Eq.~(\ref{eq:H_CR_incl_measurement_tone}))
are a deterministic force driving the mean cavity quadratures only.
As our system is linear, they thus have no impact on the noise properties
or dressed cavity susceptibility, and thus play no role in the following
discussion.

To gain an understanding of how bad cavity effects modify the measurement
rate, let us first consider a very weak measurement tone, i.e., we
focus on the limit $G_{0}\approx0$. In this case, $\hat{\tilde{H}}_{\text{CR}}\approx\hat{H}_{\text{CR}}$.
As discussed above, $\hat{H}_{\text{CR}}$ limits the maximum amount
of squeezing, cf.~Fig.~\ref{FIG_3}. However, squeezing is still
given by Eq.~(\ref{eq:squeezing_omega_0_imped_matched}) for moderate
cooperativities. Thus, the measurement rate for weak measurement tones
is still expected to scale like $e^{2r}\approx4\C$ until it is expected
to saturate to $8\Omega^{2}/\kappa^{2}$ for larger cooperativities.
Note that the assumption of a small $G_{0}$ does not necessarily
imply a weak dispersive coupling $A_{0}$. 

If we, however, were to increase the measurement tone strength further
(e.g.~to increase the absolute measurement rate $\Gamma_{\text{meas}}\propto A_{0}^{2}$),
the additional counter-rotating term $\sim G_{0}$ in Eq.~(\ref{eq:H_CR_incl_measurement_tone})
becomes more and more important. This term is expected to further
degrade the maximum achievable amount of squeezing, as the cavity
$\hat{U}_{2}$ quadrature now gets additionally coupled to $\hat{X}_{1}$.
In turn, it is expected to further limit the maximal achievable measurement
rate. Thus, the favored strategy to generate an appreciable measurement
rate $\Gamma_{\text{meas}}$ (as compared to $\Gamma_{\text{meas}}^{\text{lc}}$),
hence, would be to keep $G_{0}$ as small as possible while aiming
for a cooperativity which maximizes squeezing, and, hence, the measurement
rate.

\begin{figure}
\begin{centering}
\includegraphics[scale=1.25]{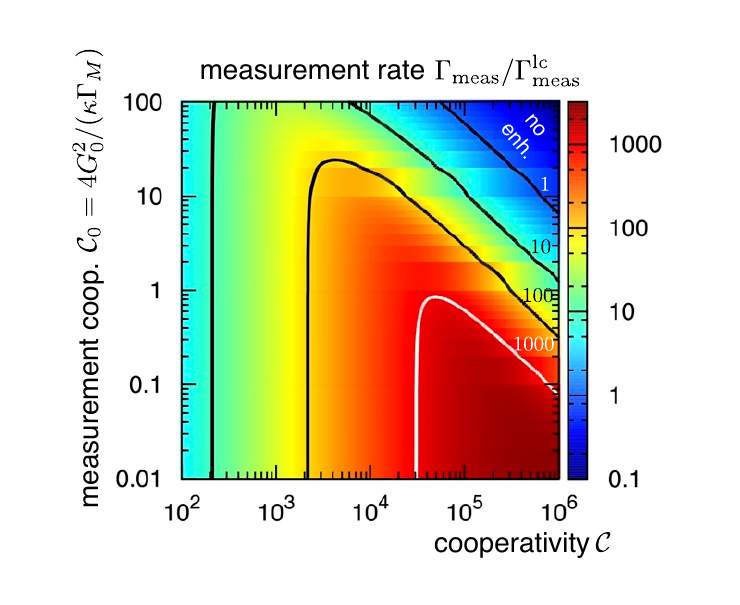}
\par\end{centering}

\caption{Enhancement $\Gamma_{\text{meas}}/\Gamma_{\text{meas}}^{\text{lc}}$
of the dispersive measurement rate by dissipative squeezing. $\Gamma_{\text{meas}}/\Gamma_{\text{meas}}^{\text{lc}}$
(cf.~Eq.~(\ref{eq:msmnt_rate_cf_linear_cavity})), i.e., the ratio
of the squeezing enhanced measurement rate to the standard measurement
rate (without optomechanical interaction), as a function of the cooperativity
$\C=4G_{-}^{2}/\left(\kappa\Gamma_{M}\right)$ and the measurement
tone driving strength (parametrized by the cooperativity $\C_{0}=4G_{0}^{2}/\left(\kappa\Gamma_{M}\right)$).
The black and white lines are contour lines depicting $\Gamma_{\text{meas}}/\Gamma_{\text{meas}}^{\text{lc}}=1,10,100,1000$.
{[}Parameters: $\Gamma_{M}=2\cdot10^{-6}\,\Omega$, $\kappa=0.05\,\Omega,$$\tilde{\kappa}=\kappa$
and $\nth=10${]}.}

\label{FIG_4}
\end{figure}

To verify our intuition, let us now focus on Fig.~\ref{FIG_4}, where
we depict the measurement rate enhancement factor $\Gamma_{\text{meas}}/\Gamma_{\text{meas}}^{\text{lc}}$
as a function of the red-laser driving strength and the measurement-tone
strength. We choose the blue driving strength $G_{+}$ to optimize
squeezing, i.e., to fulfill the impedance matching condition (\ref{eq:impedance_matching-1}).
We parametrize the red-laser strength via the cooperativity $\C=4G_{-}^{2}/(\kappa\Gamma_{M})$.
The measurement tone strength and, hence, also the strength of the
unwanted optomechanical interaction induced by the measurement tone
is parametrized via the measurement cooperativity $\C_{0}=4G_{0}^{2}/\left(\kappa\Gamma_{M}\right)$.
Note that as both $\Gamma_{\text{meas}},\Gamma_{\text{meas}}^{\text{lc}}\sim A_{0}^{2}$,
the measurement rate enhancement factor $\Gamma_{\text{meas}}/\Gamma_{\text{meas}}^{\text{lc}}$
is independent of the dispersive coupling $A_{0}$. 

For a weak measurement tone $\C_{0}\ll1$ we see that the ratio of
the measurement rates $ $ $\Gamma_{\text{meas}}/\Gamma_{\text{meas}}^{\text{lc}}$
increases linearly with the cooperativity $\C$ first until it saturates
to $\sim8\Omega^{2}/\kappa^{2}$ for large $\C$. Thus, as expected,
the unwanted optomechanical interaction induced by the measurement
tone is negligible.

Let us now increase the measurement tone strength (i.e., $\C_{0}$)
further.  For a fixed $\C_{0}$, the measurement rate enhancement
factor $\Gamma_{\text{meas}}/\Gamma_{\text{meas}}^{\text{lc}}$ exhibits
a maximum as a function of the cooperativity $\C$ as the unwanted
optomechanical interaction due to the measurement tone becomes important.
Thus, an arbitrarily large cooperativity is\textit{ not} optimal in
this regime. For realistic values of $\C_{0}$, however, we still
get a large maximum enhancement factor.

\section{Conclusion}

We have shown that strongly squeezed output light can be generated
when an optomechanical cavity is driven by two lasers on the red and
blue mechanical sideband. The output light is maximally squeezed when
an impedance matching condition (cf.~Eq.~(\ref{eq:impedance_matching-1}))
is fulfilled. Then, all incident optical vacuum fluctuations are perfectly
absorbed by the mechanical resonator and are replaced by effectively
squeezed mechanical noise.

Furthermore, we have compared our dissipative scheme to ponderomotive
squeezing and have shown that our dissipative scheme outperforms ponderomotive
squeezing in the good cavity limit which is commonly used in experiments.

We also have shown that our dissipative scheme can directly be used
to enhance the intrinsic measurement sensitivity of the optomechanical
cavity. Thus, our scheme could e.g.~be implemented in optomechanical
setups working in the microwave regime to increase the measurement
sensitivity of a dispersively coupled superconducting qubit. Note
that although we have focussed on an optomechanical implementation
of our scheme, it could also e.g.~be implemented using superconducting
circuits.

\section*{Acknowledgments}

We acknowledge support from the DARPA ORCHID program through a grant
from AFOSR, the Emmy-Noether program, the European Research Council
and the ITN cQOM. AK thanks AAC for his hospitality at McGill. 

\setcounter{section}{0}
\renewcommand{\thesection}{\Alph{section}}

\section{Appendix}

\subsection{Dissipative squeezing in the strong coupling regime\label{sub:SqueezingInTheStrongCouplingRegime}}

\begin{figure}
\begin{centering}
\includegraphics[width=0.55\textwidth]{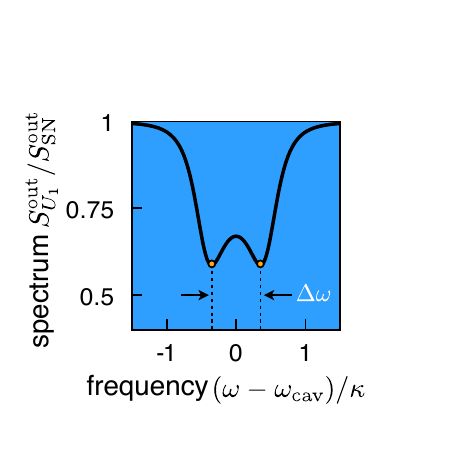}
\par\end{centering}

\caption{Output light spectrum $S_{U_{1}}^{\text{out}}$ in the {}``strong
coupling regime''. As the cavity photons and the mechanical mode
are strongly coupled, two distinct minima are observed in the output
light spectrum. Here, $\Delta\omega=2\omega_{+}=\sqrt{8\G^{2}-\kappa^{2}-\Gamma_{M}^{2}}/\sqrt{2}$.
{[}Parameters: $\Gamma_{M}/\kappa=0.1$, $\nth=10$, $r=5$ and $\G=1/2$,
i.e., $\C\approx5.5\cdot10^{3}$ and $G_{+}/G_{-}\approx1-9\cdot10^{-5}${]}.}

\label{FIG_5}
\end{figure}

In this appendix we discuss the regime where the mechanical mode and
the cavity are strongly coupled, i.e., when $\G=\sqrt{G_{-}^{2}-G_{+}^{2}}$
is appreciable. In this case, we observe a normal mode splitting in
the output light spectrum, cf.~Fig.~\ref{FIG_5}. It turns out that
squeezing is maximized at frequencies 
\[
\omega_{\pm}=\pm\sqrt{8\G^{2}-\kappa^{2}-\Gamma_{M}^{2}}/\left(2\sqrt{2}\right)\,.
\]
Thus, one enters the strong coupling regime if
\begin{equation}
8\G^{2}\ge\kappa^{2}+\Gamma_{M}^{2}\,.\label{eq:strong_coupling_condition}
\end{equation}
Note that for impedance matched parameters $\tilde{\kappa}=4\G^{2}/\Gamma_{M}=\kappa$,
the strong coupling condition (\ref{eq:strong_coupling_condition})
cannot be fulfilled. Thus, for impedance matched parameters, squeezing
is always maximized at the cavity resonance frequency. 

Let us now briefly study how the maximum achievable squeezing at $\omega_{\pm}$
depends on the damping rates $\kappa,\Gamma_{M}$ and the coupling
$\G$, i.e., we focus on the limit where the squeezing parameter $r\to\infty$.
We find
\[
\left.S_{U_{1}}^{\text{out}}/S_{\text{SN}}^{\text{out}}\right|_{\text{min}}=\frac{\left(\Gamma_{M}-\kappa\right)^{2}\left[\left(\Gamma_{M}+\kappa\right)^{2}-16\G^{2}\right]}{\left(\Gamma_{M}+\kappa\right)^{2}\left[\left(\Gamma_{M}-\kappa\right)^{2}-16\G^{2}\right]}\,.
\]
Thus, in the common limit where $\Gamma_{M}\ll\kappa$, one cannot
generate squeezing dispersively in the strong coupling regime as $\left.S_{U_{1}}^{\text{out}}/S_{\text{SN}}^{\text{out}}\right|_{\text{min}}\to1$.
If, however, $\Gamma_{M}=\kappa$, one is able to generate perfectly
squeezed output light (i.e., $S_{U_{1}}^{\text{out}}/S_{\text{SN}}^{\text{out}}|_{\text{min}}=0$)
at frequencies $\omega_{\pm}$, \textit{irrespective }of the size
of\textit{ }$\G$.

\subsection{Effects of intrinsic cavity losses and two-sided cavity}

In this appendix we focus on the dissipative generation of squeezed
output light using a single-sided optomechanical cavity in the presence
of internal losses. As we will see, internal losses will degrade the
amount of squeezing generated and the state purity. In the presence
of internal losses, the dynamics of the quantum fluctuations of the
intracavity light field reads
\[
\dot{\hat{d}}=\frac{\i}{\hbar}\left[\hat{H},\hat{d}\right]-\frac{\kappa_{\text{tot}}}{2}\hat{d}-\sqrt{\kappa_{O}}\hat{d}_{\text{in}}^{(O)}-\sqrt{\kappa_{I}}\hat{d}_{\text{in}}^{(I)}\,,
\]
where $\kappa_{\text{tot}}=\kappa_{O}+\kappa_{\text{I}}$ is the total
cavity decay rate, $\kappa_{O}$ is the photon decay rate through
the output mirror and $\kappa_{I}$ is the rate with which photons
decay internally (or e.g.~through a second, unobserved mirror). As
only the light leaving the cavity through the output mirror is of
interest, we focus on the output light described by $\hat{U}_{1}^{\text{out}}=\left(\hat{d}_{\text{in}}^{\dagger(O)}+\hat{d}_{\text{in}}^{(O)}\right)/\sqrt{2}$
where $\d_{\text{out}}^{(O)}=\d_{\text{in}}^{(O)}+\sqrt{\kappa_{O}}\hat{d}$.
For physical transparency, we assume the extreme good cavity limit,
i.e., the system's Hamiltonian is given by Eq.~(\ref{eq:Hsqueezing}).
Solving the relevant equations of motion and calculating the output
light spectrum $S_{U_{1}^{(O)}}^{\text{out}}\left[\omega\right]$
(cf.~Eq.~(\ref{eq:def_output_light_spectrum})) we find that the
output light quadrature $\hat{U}_{1}^{(0)}$ is still maximally squeezed
at the cavity frequency if the impedance matching condition $\tilde{\kappa}=4\G^{2}/\Gamma_{M}=\kappa_{\text{tot}}$
is fulfilled. The amount of squeezing at $\omega=0$ then reads
\begin{eqnarray*}
S_{U_{1}^{(o)}}^{\text{out}}/S_{\text{SN}}^{\text{out}} & = & \frac{\kappa_{I}}{\kappa_{\text{tot}}}+\frac{\kappa_{O}}{\kappa_{\text{tot}}}\left(1+2\nth\right)e^{-2r}\\
 &  & \approx\frac{\kappa_{I}}{\kappa_{\text{tot}}}+\frac{\kappa_{O}}{\kappa_{\text{tot}}}\frac{1+2\nth}{4\C}\,,
\end{eqnarray*}
where the last term is valid in the large $\C$ limit. Thus, even
if $\C\to\infty$, one cannot squeeze the output light below $\kappa_{I}/\kappa_{\text{tot}}$. 

Note that as a part of the light leaves the cavity into an unobserved
mode, the purity of the squeezed output light is not given by $\neff[0]=\nth[0]$
anymore. Instead, we find $\neff[0]\sim\sqrt{\left(1+2\nth\right)\kappa_{O}\kappa_{I}\C/\kappa^{2}}$,
i.e., the impurity increases without bound with the cooperativity
$\C$.

\subsection{Effects of laser phase noise on dissipative squeezing of light}

In this appendix we discuss the impact of laser phase noise on our
dissipative light squeezing scheme. Note that laser phase noise has
already been studied in the context of e.g., optomechanical sideband
cooling \cite{2008_Diosi_PhaseNoise_SidebandCooling,2009_Rabl_PhaseNoise_Cooling},
optomechanical entanglement \cite{2011_Abdi_PhaseNoise_Entanglement,2011_Ghobadi_PhaseNoise_Entanglement},
and back-action evasion measurement schemes \cite{2011_Phelps_PhaseNoise_BAE}. 

As before, we assume a two-tone driven optomechanical cavity, cf.~Eq.~(\ref{eq:OM_Hamiltonian}).
However, we now also take a fluctuating laser phase $\varphi(t)$
into account, i.e., the laser drive now reads 
\[
\alpha(t)=\left(\alpha_{+}e^{-\i\Omega t}+\alpha_{-}e^{\i\Omega t}\right)e^{-i\omega_{\text{cav}}t}e^{-i\varphi(t)}\,.
\]
Note that we have assumed a fixed \textit{relative} phase between
the two lasers. This implies that the maximally squeezed cavity output
quadrature is independent of the laser phase noise.

To study the impact of the (global) fluctuating phase $\varphi(t)$
on the output light squeezing, we follow the analysis of laser phase
noise presented in \cite{2011_Abdi_PhaseNoise_Entanglement,2011_Ghobadi_PhaseNoise_Entanglement}.
Thus, we go into a \textit{fluctuating} frame rotating at the fluctuating
frequency $\omega_{\text{cav}}+\dot{\varphi}(t)$, i.e., we perform
the transformation
\[
\hat{a}(t)\mapsto\hat{a}(t)\exp\left[-i\omega_{\text{cav}}t-\i\int_{0}^{t}\mathrm{d}\tau\,\dot{\varphi}(\tau)\right]\,.
\]
Note that this means that all optical quadratures have to be measured
(e.g.~in a homodyne setup) by using the same random phase noise $\varphi(t)$
as the local oscillator \cite{2011_Abdi_PhaseNoise_Entanglement}.
We now also go into an interaction picture with respect to the free
mechanical resonator Hamiltonian. Applying again standard linearization,
assuming $\dot{\varphi}\left(\bar{a}_{\pm}+\hat{d}\right)\approx\dot{\varphi}\bar{a}_{\pm}$
and applying a rotating wave approximation we finally find the equations
of motion
\begin{eqnarray*}
\dot{\hat{U}}_{1} & = & -\frac{\sqrt{2}}{g_{0}}\left(G_{-}-G_{+}\right)\dot{\varphi}\sin\Omega t-\left(G_{-}-G_{+}\right)\hat{X}_{2}-\frac{\kappa}{2}\hat{U}_{1}+\sqrt{\kappa}\hat{U}_{1}^{\text{in}}\\
\dot{\hat{X}}_{2} & = & \left(G_{+}+G_{-}\right)\hat{U}_{1}-\frac{\Gamma_{M}}{2}\hat{X}_{2}+\sqrt{\Gamma_{M}}\hat{X}_{2}^{\text{in}}
\end{eqnarray*}
and
\begin{eqnarray*}
\dot{\hat{U}}_{2} & = & \frac{\sqrt{2}}{g_{0}}\left(G_{+}+G_{-}\right)\dot{\varphi}\cos\Omega t+\left(G_{+}+G_{-}\right)\hat{X}_{1}-\frac{\kappa}{2}\hat{U}_{2}+\sqrt{\kappa}\hat{U}_{2}^{\text{in}}\\
\dot{\hat{X}}_{1} & = & -\left(G_{-}-G_{+}\right)\hat{U}_{2}-\frac{\Gamma_{M}}{2}\hat{X}_{1}+\sqrt{\Gamma_{M}}\hat{X}_{1}^{\text{in}}\,.
\end{eqnarray*}
Note that we again take $G_{\pm}$ to be real and positive, such that
the maximally squeezed cavity output quadrature is $\hat{U}_{1}^{\text{out}}$.

As dissipative light squeezing for impedance matched parameters $\tilde{\kappa}=4\left(G_{-}^{2}-G_{+}^{2}\right)/\Gamma_{M}=\kappa$
is strongest at the cavity resonance frequency (cf.~section \ref{sub:Detailed-calculation}),
we now focus on the output light spectrum at the cavity frequency
$\omega=0$. Assuming (for simplicity) a flat spectrum for the laser
phase noise, i.e., assuming $\left\langle \dot{\varphi}\left[\omega\right]\dot{\varphi}[\omega']\right\rangle =2\Gamma_{L}\delta(\omega+\omega')$
where $\Gamma_{L}$ is the laser linewidth, we find
\begin{equation}
S_{U_{1}}^{\text{out}}[\omega=0]/S_{\text{SN}}^{\text{out}}=\left(1+2\nth+\Gamma_{M}\Gamma_{L}/g_{0}^{2}\right)e^{-2r}\,.\label{eq:squeezing_omega_0_incl_phase_noise}
\end{equation}
By comparing Eq.~(\ref{eq:squeezing_omega_0_incl_phase_noise}) to
our previous finding (\ref{eq:squeezing_omega_0_imped_matched}),
we see that (global) laser phase noise effectively increases the mechanical
bath temperature only. Thus, (global) phase noise is negligible if
\[
\Gamma_{L}\ll g_{0}^{2}/\Gamma_{M}\,.
\]
Note that this condition is equivalent to the one found in Ref.~\cite{2009_Rabl_PhaseNoise_Cooling}
which has to be fulfilled to be able to achieve optomechanical ground
state cooling. 

Thus, we conclude that (global) phase noise should not pose a strong
limitation on our dissipative squeezing scheme.

\section*{References}

\bibliographystyle{iopart-num}
\bibliography{Optomechanics}

\end{document}